\documentstyle[prl,aps,epsf]{revtex}
\draft
\begin{document}
\twocolumn[\hsize\textwidth\columnwidth\hsize\csname @twocolumnfalse\endcsname

\title{Charge Transport Transitions and Scaling in Disordered Arrays
of Metallic Dots
} 
\author{C. Reichhardt and C.J. Olson Reichhardt} 
\address{ 
Center for Nonlinear Studies and Theoretical Division, 
Los Alamos National Laboratory, Los Alamos, New Mexico 87545}

\date{\today}
\maketitle
\begin{abstract}
We examine the charge transport through disordered arrays
of metallic dots using numerical simulations. 
We find power law scaling in the current-voltage curves 
for arrays containing no voids, while for void-filled arrays 
charge bottlenecks form and a single 
scaling is absent,
in agreement with recent experiments.
In the void-free case we also show that the scaling 
exponent depends on the effective dimensionality of the system.  
For increasing applied drives we find a transition from 
2D disordered filamentary flow near threshold to a 1D smectic 
flow which can be identified experimentally 
using characteristics in the transport curves and conduction noise. 
\end{abstract}
\vspace{-0.1in}
\vspace{-0.2in}

\vskip2pc]
\narrowtext
A wide variety of disordered systems exhibit threshold behavior and nonlinear
response to an applied drive. 
Examples include flux lines in disordered superconductors, 
\cite{Jensen1,Higgins2,Dominguez3} 
charge-density-waves (CDW's) pinned by impurities \cite{Grunner4}, 
Wigner crystals \cite{Williams5,Wigner6}  
in semiconductors with charge impurities, and
colloids flowing over rough surfaces \cite{Ling7}. 
Another example is charge 
transport through metallic dot arrays. 
Middleton and Wingreen (MW) have considered
a model of this system in which the randomly charged dots 
are separated by tunnel barriers \cite{Wingreen8}. 
They found threshold behaviors and scaling of the current-voltage curves
of the form $I = (V/V_{T} - 1)^\zeta$. 
In 1D they obtain $\zeta=1.0$, while
for 2D they predict analytically
$\zeta = 5/3$ and find in simulations $\zeta = 2$.
For the 2D systems the simulated 
flow patterns are not 
straight but form intricate meandering paths with considerable
transverse fluctuations. 
These same types 
of meandering paths are also observed in the 
flow of flux-lines \cite{Jensen1,Dominguez3},
Wigner crystals \cite{Wigner6}, and colloids \cite{Ling7} above the
depinning threshold.  
Experimental studies in metal dot arrays have also found 
scaling in the I-V curves for 2D and 1D systems 
\cite{Rimberg9,Duroz10,Bezyryadin11,Black12,Kurdack13};
however, the scaling exponents in these experiments exhibit a wide range of 
values. The  studies for 1D arrays \cite{Rimberg9} find a scaling 
exponent of $\zeta = 1.36$, which is less than the 
value for 2D arrays predicted by MW, but still larger than 
the expected 1D value of $\zeta=1.0$. 
It is not known if this system  
is truly 1D, or whether some meandering of the 
charge in 2D can still occur due to the finite width of the dots.
It is also not known how the exponents would change (or whether
there would even be scaling) upon changing the system from 2D to 1D 
by gradually narrowing the array width. 
Other systems in 2D exhibiting scaling near
depinning also show a wide spectrum  
of scaling exponents \cite{Jensen1,Higgins2,Dominguez3,Wigner6,Ling7},  
suggesting that the type of disorder and the effective 
dimensionality of the  
array plays a crucial role in the transport.      

Recently, to address the role of different types of disorder on transport, 
Parthasarathy {\it et al.} \cite{Jaeger14} 
performed experiments
on triangular monolayers of gold nanocrystals.
Disorder is present      
in the form of  charge disorder in the substrate as well as
variations in the interparticle couplings.  Structural disorder 
was also introduced by creating 
voids in the arrays. 
The void-free arrays exhibit robust power law scaling with
$\zeta = 2.25$.  
A single power law could {\it not} be fit for
the structurally disordered arrays. 
Parthasarathy {\it et al.} conjecture that in
arrays with voids, the charge must be shuttled into bottleneck
regions, reducing the amount of charge flow. 
In the theoretical studies of MW where
structural disorder was not considered, only power law behavior
was observed \cite{Wingreen8}.

Since the charge flow at depinning 
in the dot arrays resembles
that seen in other systems such as flux lattices,
it is of interest to
ask whether some of the ideas developed in these other systems can be carried
over to the metallic dot system. In disordered vortex systems,
theory \cite{Koshelev15,Balents16,Giamarchi17}, 
simulation \cite{Koshelev15,Olson18}, and 
experiment \cite{Pardo19} show that 
at low drives the vortex flow is highly disordered, 
meanders in 2D, and the overall  
lattice structure is destroyed. 
For high drives there can be 
a remarkable reordering transition 
where the flow occurs in 1D channels and the lattice regains 
considerable order. For strong quenched disorder in 2D,
the highly driven phase forms a moving smectic
with order in the transverse but not the longitudinal direction. 
Here the vortices
move in well spaced channels, and the channels are decoupled from 
one another. 
   
In order to compare to the recent experiments and to explore the
points raised above, 
we conduct simulations of
a simple model for charge transport in 2D arrays 
with charge disorder and both with and without structural order. 
In our model we consider square and rectangular arrays of side $N\times M$, 
with periodic boundary conditions in the
$x$ and, for the 2D systems, 
$y$ directions, containing $N_c$ mobile charges. 
Under an applied drive $f_{d}$ 
a mobile charge on a site experiences a maximum threshold force 
$f_{th}$ before exiting the plaquette. 
For actual dot arrays the applied drive comes from an applied voltage
$V$, and the energy to add an electron to a dot with charge $q$ is 
$V_{th} = q/C$ where $C$ is the capacitance of the dot. 
Charge flow will then occur for applied drives $V > V_{th}$ for a 
single dot. In addition, the mobile charges interact via
a  Coulomb term $U = q^{2}/r$. Since the Coulomb term is long range 
we use a summation technique \cite{Jenesen20} for numerical

\begin{figure}
\center{
\epsfxsize=3.5in
\epsfbox{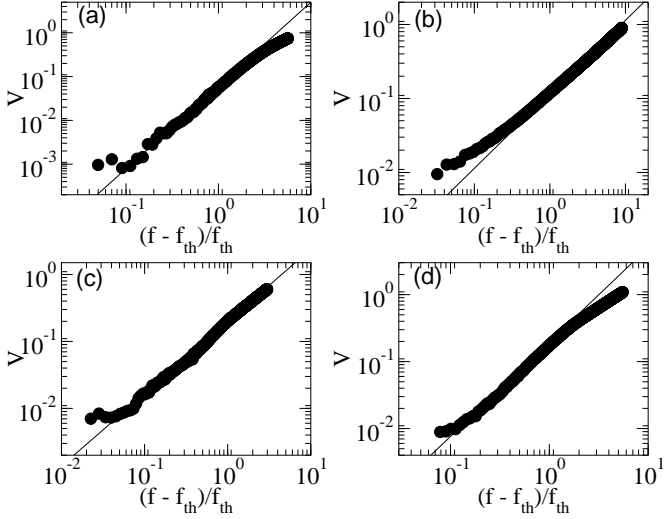}}
\caption{
The scaling of the average velocity $V$ vs applied drive f for 
(a) a 2D system with a fit of $\zeta = 1.9$
(solid line) and (b) a 1D system 
with a  $\zeta = 1.0$ fit (solid line). 
$f_{th}$ is the threshold force at which depinning occurs.
(c) A system of size $ 2 \times 500$, showing 
a fit with $\zeta = 1.0$. (d) A system of size $5 \times 500$,
with a $\zeta = 1.41$ fit.  
}
\end{figure}

\hspace{-13pt}
efficiency. For arrays without voids, we add disorder by 
selecting random thresholds $V_{th}$ from 
a Gaussian distribution centered at $V^{0}_{th}$.
To study the effect of structural disorder, a portion
$P < 0.5$ of randomly 
selected sites is effectively voided so that mobile charge cannot
flow into them. 
We do not consider thermal effects since our 
system is in the Coulomb-blockade regime, so that charging 
energies are higher than the thermal energies.
For increasing applied drive or voltage we measure the global charge flow
or current $I$. We also measure the trajectories of the flow and 
the fluctuations in the current, which is
proportional to the conduction noise.

We first consider the scaling for ordered arrays in 2D, 1D, and
finite width samples. 
In Fig.~1(a) we show the scaling of the average flow vs applied
drive (I-V) curve for the 2D case, and in Fig.~1(b) the 
1D case. 
The curves in Fig.~1(a,b) are for the largest systems we have considered;
for smaller systems the scaling regime is reduced.
For 2D we find scaling with $\zeta = 1.9 \pm .15$ 
in fair agreement with the simulation results $\zeta = 2.0$ of MW, but still 
lower than the $\zeta=2.25$ found in Ref.~\cite{Jaeger14} for ordered arrays.
The simulations for 1D 
give a linear behavior
for much of the curve,
suggesting that if a scaling exponent could be 
ascribed it would be $\zeta < 1.0$; however, for drives near 
threshold, a single scaling cannot be applied and the curve
bends up.
We note that for the depinning of 
1D elastic objects such as
CDW's one expects
an exponent of $\zeta = 1/2$ \cite{Grunner4}. 
Experiments measuring the threshold
of a {\it single} dot also find $\zeta = 1/2$ \cite{Duroz10}.
Fig.~1(b) shows that the deviation from linearity occurs only
very near threshold, so that the discrepancy between our results and MW
may result from MW not being close enough to the

\begin{figure}
\center{
\epsfxsize=1.8in
\epsfbox{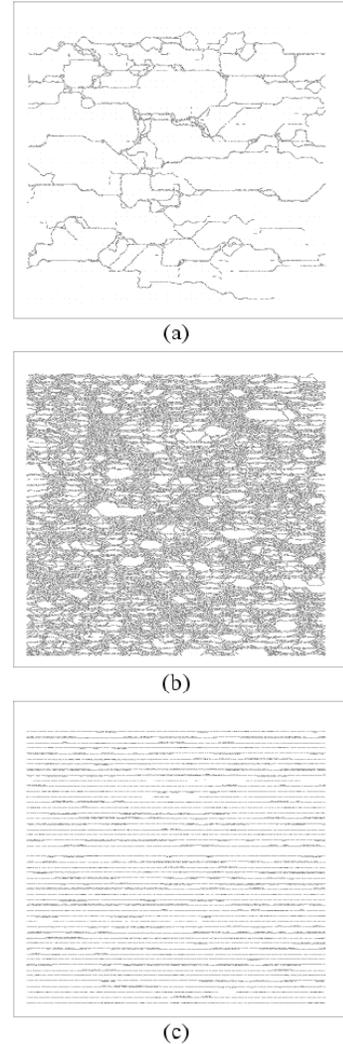}}
\caption{
The current paths for increasing applied drives in the 2D
system shown in Fig. 1(a). 
Here (a) $f_{d}/f_{th} = 1.1$ and (b) $f_{d}/f_{th} = 2.0$ produce 
meandering 2D channels. In (c) $f_{d}/f_{th} = 10.0$ produces 
straight 1D channels. 
}
\end{figure}

\hspace{-13pt}
threshold to see the
deviation.
In Fig.~1(c) we show the scaling for a system with a finite width of
$2\times 500$, with $\zeta=1.0$, and
in Fig.~1(d) the scaling for a wider system of $5\times 500$ is shown, 
where we find scaling with $\zeta = 1.41$.    
We note that experiments for the 1D arrays \cite{Duroz10} find $\zeta = 1.36$. 
Our results suggest that scaling can occur for systems
between 1 and 2 dimensions with the value of the exponent
monotonically increasing
from $1/2$ in 1D to 2.0 in 2D. Additional evidence for the
increase of the exponent as a function of dimensionality has been 
obtained in colbalt nanocrystal samples of finite
thickness, with an effective dimensionality between 2 and 3.
Here 
exponents of $2.2 < \zeta < 2.7$ are observed \cite{Black12}. 
Our results also suggest that the experiment in \cite{Jaeger14} is 
in an effective dimension higher than 2.0.   
All of the curves in Fig.~1 show a crossover to a linear regime at 
high drives, 
which was also seen in the earlier numerical work.
We do not find any hysteresis for either the 2D or 1D case. 

\begin{figure}
\center{
\epsfxsize=3.5in
\epsfbox{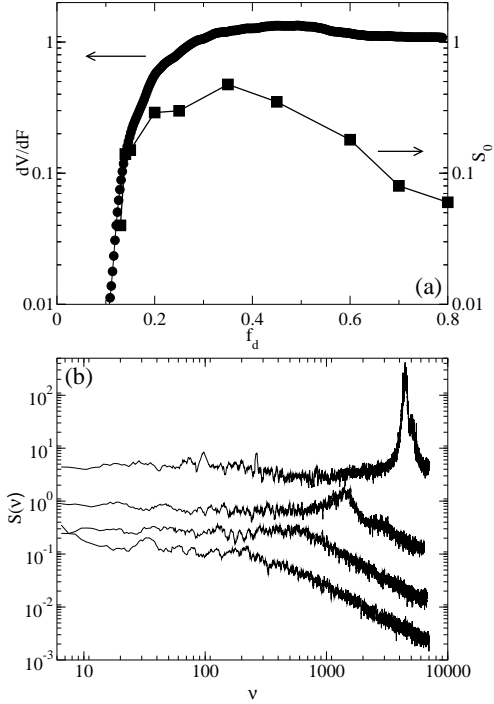}}
\caption{
(a) Circles: $dV/df$ of the curve shown in Fig.~1(a) for a 2D system. 
Squares: The
noise power $S_{0}$ as a function of drive. 
(b) Consecutive power spectra for increasing $f_{d}$ for, 
from bottom to top, $f_{d}/f_{th} = 1.1$, 2.0, 5.0, and 8.0. 
The curves have been shifted up 
for presentation. 
}
\end{figure}

We next consider the flow transitions for increasing applied drive
in the 2D system.  
In Fig.~2 we show the
current paths for three different applied drives above threshold.
In Fig.~2(a), at $f_{d}/f_{th} = 1.1$ the flow follows meandering paths in
only a few regions of the sample, in agreement with 
the simulations of MW \cite{Wingreen8}. 
These flow paths are {\it stationary} over time. 
For higher drives, as seen in Fig.~2(b) at $f_{d}/f_{th} = 2.0$, 
there is a crossover from the static paths to {\it dynamic} paths which
open, close, and shift position over time.  The 
filamentary flow in this case occurs 
everywhere in the system over time.
For drives at and above the $f_{d}$ value where the I-V curve becomes 
linear, 
as seen in Fig.~2(c) for $f_{d}/f_{th} = 10.0$, there
is a crossover from the meandering 2D flow 
to straight 1D {\it ordered} channels of flow. 
Fig.~2(c) shows that the charge moves only in 
1D channels without any 
jumping of charge between adjacent channels. The 
channels themselves carry different amounts of flowing charge due 
to the different average disorder along the rows. 
The charges in one channel do not synchronize 
with the flow of charge in adjacent channels;
instead, the channels {\it slide} past one another. 
We term this a smectic flow state since the channels are periodically spaced
in the transverse direction but are independently
moving in the longitudinal direction. 
The transition from the disordered to partially ordered flow is very 
similar to the reordering transitions seen for 
driven vortex lattices 
\cite{Koshelev15,Balents16,Giamarchi17,Olson18,Pardo19}. 
We observe the same transitions in the systems of 
finite widths. 

In Fig.~3(a) we plot the $dV/df$ curve, which is
proportional to the resistance $R$,
for the 2D system in Fig.~1(a).
The crossover to the linear 
regime (with constant $R$) appears as the plateau region in $dV/df$. 
Also shown in
Fig.~3(a) is the power $S_{0}$ from one octave of the
power spectra of the conduction noise at four different
applied drives $f_{d}$:
$S_{0}=\int_{\nu_{1}}^{\nu_{2}} d\nu S(\nu)$,
where
$S(\nu) = |\int V_{x}(t)e^{-i2\pi \nu t}dt|^2$. 
The noise power shows a peak in the scaling regime
near $f_{d}=0.35$ and then decreases
in the linear regime with a low value of $S_{0}$ during the smectic flow.
In Fig.~3(b) we show that a clearer signature of the flow phases can
be obtained by examining 
individual power spectra. For drives in the scaling
regime we find a $1/f^{\alpha}$ power spectra, which is indicative of the
many different frequencies generated by the complex flow patterns
illustrated in Fig.~2(a,b). The large noise power and $1/f^{\alpha}$ signals 
have also been
associated with meandering disordered flow in superconducting vortices. 
For increasing 
drive a characteristic peak in the spectra begins to to appear and is 
most prominent in the smectic flow regime.
This peak occurs when the charge in the smectic phase flows 
in 1D paths along the dots which 
are in a periodic array of spacing $a$. 
The frequency at which the peak occurs
is then $\nu = v/a$, where $v$ is the average velocity of the charge.
For perfectly ordered flow, the peaks would be very narrow. Since the 
channels have different amounts of flowing charge there is some 
dispersion in the frequency. 
The presence of peaks in the power spectra suggests that it 
would be possible to 
observe an interference effect, or Shapiro steps, in the I-V curves if
an additional applied AC drive is imposed. 
The steps occur when the frequency of the
AC drive matches the internal frequency of the system.

Another experimental probe of the moving smectic phase is the presence of 
a {\it transverse depinning barrier} 
as first predicted in \cite{Giamarchi17} for
elastic media. If a transverse force is applied  to the already
longitudinally moving system, then in the disordered regime there is
no threshold for transverse motion and some charge will immediately 
begin to move in the transverse direction.
For the high drive regime, after the 1D moving channels have formed,
there is a finite transverse threshold since the channels are effectively
pinned in the transverse direction. 
There may also be interesting results for
driving along different directions of the  dot lattice. Although
there is randomness in the individual dot strength, the overall 
topological order of the array can break the symmetry so that certain
directions may allow easier charge flow than 
others.

We next consider the structurally disordered arrays. We find that for a fixed
applied drive, the current is reduced as the void fraction $P$ is 
increased. This is 
understandable since the charges must 
flow in increasingly winding patterns to pass the voids,
which are effective
obstacles. In  Fig.~4(a) we plot $V$ vs $P$ up to $P=0.49$ 

\begin{figure}
\center{
\epsfxsize=3.5in
\epsfbox{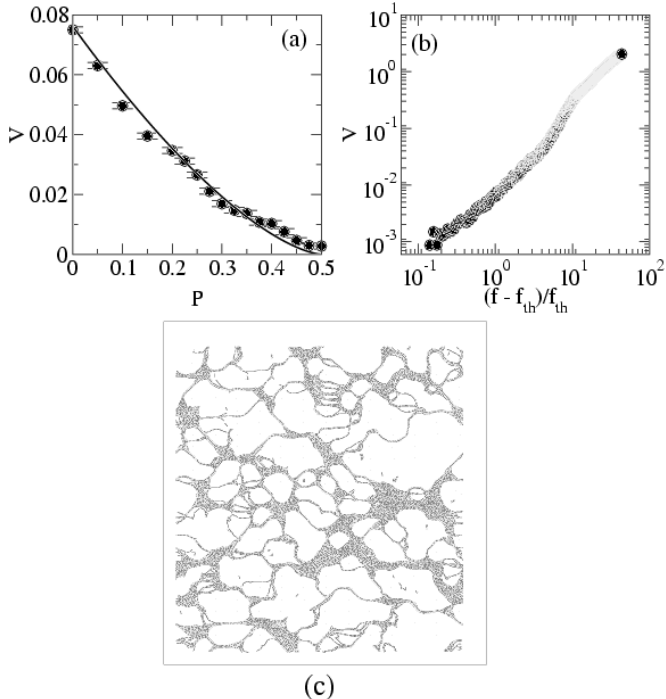}}
\caption{  
(a) The average velocity at a fixed $f_{d}=0.2$ for a system with increasing
void fraction $P$. The solid line is a fit with $V = 2(0.5 - P)^{1.5}$.
(b) The velocity vs applied drive curve for a void fraction of $P=0.47$.
(c) The trajectories for a system with a void fraction of $P=0.47$ at 
$f_{d}/f_{th} = 3.0.$  
}
\end{figure}

\hspace{-13pt}
for a fixed $f_{d} = 0.2$.  For the decreasing value of $V$, we find
a best fit to a power law with $V = V_{0}(0.5 -P)^{1.5}$. The
current should go to zero for $P = 0.5$, or the percolation limit, where 
for an infinite system voids would span the entire system. In our 
system there is still transport for $P > 0.5$ due to finite size effects.  
In Fig.~4(b) we show the current vs applied drive curve for 
a void fraction of $P=0.47$. In this case,
we {\it cannot} fit a single power law above threshold. 
In Ref.~\cite{Jaeger14} 
the additional features in the I-V curves for the
structurally disordered arrays were conjectured to
occur due to bottleneck effects caused by the void regions. 
In Fig.~4(c) we illustrate the current paths for $f_{d}/f_{th} = 3.0$
and $P=0.47$,
showing that there is considerable flow through the system, 
but that in certain  well-defined areas 
no flow is occurring.  Averaging the trajectories over a longer time
produces the same flow patterns shown in Fig.~4(c). 
This is in contrast to the
flow pattern in Fig.~2(b), which changes over time, so that for 
long times 
flow occurs in all regions of the sample.  
In Fig.~4(c) some bottlenecks can also be seen in the form of
regions where the trajectories are compressed. 
For the structurally disordered arrays the transition to the smectic flow 
at higher drives 
is absent since straight 1D flows cannot occur even in the ohmic regime.
The conduction noise shows the same $1/f^{\alpha}$ behavior as the
ordered arrays but the peaks in the noise spectra are absent in the ohmic 
regime.    

In summary, we have investigated charge transport in
structurally ordered and 
disordered arrays. For ordered arrays we find scaling in the current vs 
applied drive curves with $ \zeta = 1.9$ in 2D. For 1D arrays  
we find that the scaling exponent is less than one. 
Scaling still occurs for systems with finite 
width, with the exponent increasing toward $\zeta=2.0$ for  
increasing sample widths. 
For increasing applied 
drive in 2D, we show that the crossover to ohmic behavior coincides
with a change in the flow from 2D meandering to straight 1D channels
or smectic flow. Evidence for this change in the flow also 
appears in the form of 
a crossover 
in the power spectra, which shows a broad 
$1/f^{\alpha}$ signature
in the disordered flow regime, and a characteristic peak or washboard
signal in the smectic flow regime. 
For disordered arrays where a fraction of the sites  
are replaced with voids, a single power law cannot be fit 
to the I-V curve in agreement with 
recent experiments. The transition from the 2D disordered
flow to the 1D channel
flow is absent in the structurally disordered arrays.  

We thank A.A. Middleton, H. Jaeger, and R. Parthasarathy 
for useful discussions. 
This work was supported by the U.S. Department of Energy
under Contract No. W-7405-ENG-36.


\vspace{-0.2in}
\begin{references}
\vspace{-0.5in}

\bibitem{Jensen1}
H.J.~Jensen, A.~Brass, Y.~Brechet, and A.J.~Berlinsky,
Phys.~Rev.~B {\bf 38}, 9235 (1988). 

\bibitem{Higgins2}
S.~Bhattacharya and M.J.~Higgins, Phys.~Rev.~Lett.~{\bf 70}, 2617 (1993);
Phys.~Rev.~B {\bf 49}, 10005 (1994); 
M.~Danckwerts, A.R.~Goni, and C.~Thomsen, {\it ibid.}~{\bf 59}, R6624 (1999). 

\bibitem{Dominguez3}
D.~Dom{\' \i}nguez, Phys.~Rev.~Lett.~{\bf 72}, 3096 (1994).

\bibitem{Grunner4}
G.~Gr{\" u}nner, Rev.~Mod.~Phys.~{\bf 60}, 1129 (1988).

\bibitem{Williams5}
F.I.B.~Williams {\it et al.}, Phys.~Rev.~Lett.~{\bf 66}, 3285 (1991). 

\bibitem{Wigner6}
C.~Reichhardt {\it et al.}, Phys.~Rev.~Lett.~{\bf 86}, 4354 (2001). 

\bibitem{Ling7}
A.~Pertsinidis and X.S.~Ling 
(unpublished).  

\bibitem{Wingreen8}
A.A.~Middleton and N.S.~Wingreen, Phys.~Rev.~Lett.~{\bf 71}, 3198 (1993).

\bibitem{Rimberg9}
A.J.~Rimberg {\it et al}, Phys.~Rev.~Lett.~{\bf 74}, 4714 (1995). 

\bibitem{Duroz10}
C.I.~Duruoz {\it et al.}, Phys.~Rev.~Lett.~{\bf 74}, 3237 (1995). 

\bibitem{Bezyryadin11}
M.N.~Wybourne {\it et al.}, Jpn.~J.~Appl.~Phys.~{\bf 36}, 7796 (1997);
A.~Bezryadin {\it et al.}, Appl.~Phys.~Lett.~{\bf 74}, 2699 (1999).

\bibitem{Black12}
C.T.~Black {\it et al.}, Science {\bf 290}, 1131 (2000).

\bibitem{Kurdack13}
C.~Kurdak {\it et al.}, Phys.~Rev.~B {\bf 57}, R6842 (1998).

\bibitem{Jaeger14}
R.~Parthasarathy, X.-M. Lin, and H.M. Jaeger, Phys. Rev.~Lett.~{\bf 87}, 186807
(2001).

\bibitem{Koshelev15}
A.E.~Koshelev and V.M.~Vinokur, Phys.~Rev.~Lett.~{\bf 73}, 3580 (1994).

\bibitem{Balents16}
L.~Balents, M.C.~Marchetti, and L.~Radzihovsky,
Phys.~Rev.~Lett.~{\bf 78}, 751 (1997);
Phys.~Rev.~B {\bf 57}, 7705 (1998). 

\bibitem{Giamarchi17}
P.~Le Doussal and T.~Giamarchi, Phys.~Rev.~B {\bf 57}, 11 356 (1998).

\bibitem{Olson18} 
K.~Moon {\it et al.}, Phys.~Rev.~Lett.~{\bf 77}, 2778 (1996);
C.J.~Olson, C.~Reichhardt, and F.~Nori, {\it ibid.}~{\bf 81}, 3757 (1998);
A.B.~Kolton {\it et al.}, {\it ibid.}~{\bf 83}, 3061 (1999). 

\bibitem{Pardo19}
F.~Pardo {\it et al.}, Nature (London) {\bf 396}, 348 (1998).

\bibitem{Jenesen20}
N.~Gr{\o}nbech-Jensen, Int. J. Mod. Phys. C {\bf 8}, 1287 (1997). 

\end{references}
\end{document}